\documentclass[apjl]{emulateapj}
\usepackage{epsfig}

\begin{document}

\title{The Warped Circumstellar Disk of HD100546
}

\author{Alice C. Quillen}
\affil{Department of Physics and Astronomy, University of Rochester, Rochester, NY 14627}
\affil{Visitor, Research School of Astronomy and Astrophysics, 
Australian National University, 
Mount Stromlo Observatory, Cotter Road, Weston Creek, ACT 2611, Australia
}
\email{aquillen@pas.rochester.edu}

\begin{abstract}
We propose that the two armed spiral features seen in visible {\it Hubble
Space Telescope} images
of scattered light in HD100546's circumstellar disk 
are caused by the illumination of a warped outer disk.
A tilt of 6-15 degrees from the symmetry plane can cause the observed
surface brightness variations providing the disk is 
very twisted (highly warped) 
at radii greater than 200 AU where the spiral features are seen.  
Dust lanes are due in part to shadowing in the equatorial plane from the
inner disk within a radius of 100 AU.
HD100546's outer disk, if viewed edge-on, would appear similar 
to that of Beta Pictorus.
A disk initially misaligned with a planetary system,
becomes warped due to precession induced by planetesimal
bodies and planets.
However, the twistedness of HD100546's disk cannot be explained 
by precession during the lifetime of the system 
induced by a single Jovian mass planet within the clearing at $\sim 13$ AU.
One possible explanation for the corrugated disk is that
precession was induced by massive of bodies embedded in the disk at larger
radius.  This would require approximately a Jupiter mass of bodies 
well outside the central clearing at 13 AU and
within the location of the spiral features 
or at radii approximately between 50-200 AU. 
\end{abstract}


\section{Introduction}

High angular resolution imaging of circumstellar disks
have shown that these disks may not be coplanar. 
For example, the disk of Beta Pictorus exhibits a warp; the
outer disk at radii greater than 50AU 
is tilted between 3-5 degrees from the disk interior
\citep{burrows,heap}.  Models for the scattered optical light
of the bowed disk of AU Microscopii suggest that this disk is also warped, 
with a small tilt of about $3^\circ$ \citep{krist05}.
While warped disks are easiest to identify in nearly edge-on
systems such as Beta Pictorus and AU Microscopii, 
less inclined or nearly face-on disks may also not be coplanar.
This leads us to search for features in less inclined disks
that might also be explained with a warped disk model.
As proposed by \citet{mouillet97, augereau01b},
the twistedness of Beta Pictorus's warped disk could be a result of precession
induced by unseen planets or planetesimals residing in the disk.
Consequently by probing the geometry of the warp, we can constrain
the properties of the unseen planetary system.

HD100546 is a nearby southern Herbig Be star (B9.5Ve; distance $d=103\pm 6$pc
with an age estimated $t\sim 10$Myrs  \citep{vandenancker} which 
exhibits a dusty circumstellar disk revealed in scattered light
from visible and near-infrared imaging \citep{grady,augereau01}.
We note that at the distance of HD 100546, $1''$ corresponds
approximately to 100AU.
Most noticeable in the images of this disk taken by the {\it Hubble
Space Telescope (HST)} using the {\it STIS} camera
are the two nearly symmetrical spiral features at a
radius of about
250 AU from the star \citep{grady}.   \citet{quillenhd} showed
that the two arms 
were unlikely to be caused by spiral density waves excited by
a planet embedded in the disk but   
could be density variations pulled out tidally by a nearby passing star.
However, \citet{quillenhd} then ruled out this scenario 
for a number of reasons. 
To match the openness of the arms, the flyby had to be recent;
within a few thousand years.  Not only is  the probability for this 
scenario extremely low, but it implies that the perturbing object
is near HD100546.  A search for a perturbing object (star) massive
enough to account for the two arms was negative.
\citet{quillenhd} concluded that the spiral features were not 
density enhancements resulting from the tidal force of a recent stellar
close encounter.  
In contrast, both \citet{quillenhd} and \citet{ardila} 
have shown that the spiral arms in the outer part of HD141569A's disk
{\it are} likely to haven been pulled out tidally by the nearby passage of
it's binary companion, HD141569B,C.

Warped disks can display complex morphology.
Previous work modeling galactic warped disks 
(e.g., \citealt{quillenCO,quillen,quillenm84, nicholson, tubbs,quillenircena,steiman92})
accounts for the optical dust lanes seen in Centaurus A 
and other galaxies and the parallelogram
shaped infrared morphology seen in Spitzer images of Centaurus A.
These previous studies have shown that the apparent morphology
of a warped disk can be sensitive to its optical properties 
(e.g., \citealt{krist05})
as well as its geometry.  For emitting optically thin disks,
folds in the disk can appear brighter to the viewer than other regions
(see \citealt{quillenircena} and associated simulations).
Likewise we expect that optically thin disks appear brighter
in scattered light
in regions where the disk tilts at high inclination with respect to
the viewer.  This is because the surface brightness 
of an optically thin medium of uniform thickness
that scatters light from a distant source
is dependent on its orientation angle with respect to the viewer. 

The outer disk is illuminated by starlight that may pass through
an inner absorbing disk.  Consequently starlight
illuminating the outer disk should be 
attenuated in the equatorial plane.  
A warped disk can be described as a series of rings.  
Each ring crosses the equatorial plane at two points.
The shadow from  the inner disk could 
cause two spiral dustlanes or shadows in 
the outer warped disk, resulting in
the appearance of two spiral features.

Here we consider the possibility that the disk of 
HD100546 is warped, and that the warp is responsible for the observed
spiral features.  We first consider the case
where the outer disk is isotropically illuminated by the central star.
If the disk at large radii 
is warped and optically thick then
we would expect the optical images would exhibit only one spiral arm. 
Only the high side of the disk would be illuminated whereas 
the opposite side would be in shadow. 
If the disk is optically thin and warped at radii greater than $2''$ 
or 200AU, then we expect
to see two spiral features.   Regions of high inclination with respect
to the viewer would appear brighter and would be located on opposite
sides of the star.  

We now consider the situation of a warped disk which is not isotropically
illuminated.
We expect that starlight illuminating the outer disk should be highly 
attenuated in the equatorial plane.
At each radius an inclined disk crosses the equatorial plane
at two opposing points.
Consequently the shadow from  the inner disk could 
cause two spiral dustlanes or shadows in 
the outer warped disk, resulting in
the appearance of two spiral features.
If the spiral features are due to an equatorial shadow then 
the outer disk could either be optically thin or thick.
If the disk is optically thick then one would expect 2 spiral shadows
and one bright spiral feature, wheres if the disk is optically thin
then one would expect 2 spiral shadows with brighter spiral features
lying between them.

Unfortunately it is not straightforward to estimate
the disk optical depth and width of HD100546's disk as a function
of radius.  
Based on the near-IR scattered light surface brightness, 
\citet{pantin} estimated that the disk
has a high normal optical depth, $\tau \sim 1$, at 
radii greater than $1''$ or 100AU.
However, \citet{augereau01} estimated a normal optical depth 40-50 times
lower implying that the disk was optically thin at optical 
and near-IR wavelengths outside of $\sim 80$ AU.
The observed infrared spectral energy distribution 
of HD100546 implies that the luminosity of emission from dust 
is large, 50\% of the stellar photospheric luminosity.  
Most of the disk luminosity, 70\%, is emitted in the mid-infrared 
and so from a radius $\sim 10-20$ AU \citep{bouwman}, well
within the location of the spiral features at $\sim 250$ AU in
scattered light. 
The estimated radius of the mid-infrared emission based
on the spectral energy distribution
is approximately consistent with 
mid-infrared high angular resolution observations
\citep{liu03}.
The infrared spectral energy distribution can be explained with a disk 
containing an inner, puffed up region absorbing and re-radiating
half of the stellar light, accounting
for the large mid-infrared flux. Because
of its large covering angle the puffed inner disk edge shades the outer disk
\citep{bouwman,dullemond01,dullemond04}.
However, this model
implies that the disk greater than $1''$ or 100AU which
has been observed in scattered light 
is illuminated by the starlight that must pass first through
the inner puffed up region of the disk.  
This situation makes it challenging
to estimate the optical depth of the disk
from the scattered light at radii greater than $1''$.

In this paper we search for 
warped disk models which can account
for the spiral features seen in HD100546's disk.
Warped disk models have some advantages over other models.  
Because planetary systems are nearly Keplerian, 
orbit orientations vary extremely slowly. 
Consequently a warp 
can be maintained for many rotation periods.  This eliminates
one of the problems of the previously proposed transient spiral 
structure model.
Edge-on systems such as Beta Pictorus which are clearly warped,
suggest that less highly inclined systems such as HD100546 
could also be warped.
In section 2 we describe how
we represent the geometry of a warped disk. 
We then explain how we synthesize model surface brightness images which
can be compared to the observations.
In section 3 we discuss dynamical or physical models for the disk warp based
on the geometrical model which best matches the observed disk morphology.
A summary and discussion follows.

\section{Warped disk models}

\subsection{Warp geometry}

We first discuss our notation for describing the orientation of
the planetary system with respect to the viewer.
The orientation of a coplanar planetary system 
requires 2 angles to describe; $\chi$, corresponding to the position angle
(counter clockwise from North) of the axis of disk rotation on the sky, 
and an inclination angle, $\vartheta$, which describes the tilt of this axis
with respect to the line of sight.  If $\vartheta$ is zero, then
the planetary system is viewed face-on.
For a system that is not coplanar, 
$\chi$ and $\vartheta$ refer to the orientation of the rotation
axis corresponding to the total angular momentum of the system's 
disk and planets.

We describe the warped disk with respect to the rotation
axis of the planetary system.
A warped disk undergoing circular motion
can be described as a series of rotating tilted
rings, each with a different radius, $r$.  
The orientation of each ring is specified by two angles, 
a precession angle, $\Omega(r)$, similar
to the longitude of the accenting node, and an inclination angle, $i(r)$.  
These angles are given
with respect to the rotation axis of the planetary system 
and the direction of the line of sight. 
We measure the angle $\Omega$ 
from a reference line in the system's ecliptic.  
As viewed on the sky,
this line lies on the rotation axis of the system, 
but is projected onto the system's ecliptic plane.
Our precession angle is the angle (measured at the star)
between this line 
and the point at which the ring crosses the system's ecliptic.
This is similar to the longitude of the ascending node which
is measured with respect to the vernal equinox and the point
at which the orbit crosses the ecliptic.

\subsection{Constructing a model scattered light image}

To produce an model image of the optical scattering light, 
all reflecting and absorbing regions along the line of sight 
at each position on the sky 
must be considered. When the disk is optically thin, multiple scattering
events and absorption can be neglected.  
In this case each photon originates from the star and is
then reflected from a single spot on the disk.
At each position on the sky  
we can sum the reflected light at each location in the disk
along the line of sight.
We have restricted this modeling effort to an optically thin outer disk,
consistent with the estimates by \citet{augereau01} at
radii outside of 80AU.  However we must
keep in mind that future modeling efforts may need to consider outer
disks with higher optical depth.

We begin by randomly sampling $x,y$ positions in the plane perpendicular
to the disk rotation axis.  At each
position we compute a disk plane $z$ coordinate based on our assumed 
function for the precession and inclination angles 
$\Omega(r)$ and $i(r)$.  To account for
the disk thickness we add a vertical offset to $z$
which was randomly chosen
from a Gaussian distribution function.
For the vertical structure of the disk we assume a normal distribution
\begin{equation}
\rho(z) \propto \exp(-z^2/2 h^2)
\end{equation}
where $h$ is the standard deviation of the distribution.
The FWHM of this distribution is $2.35h$.
We adopt a disk aspect ratio, $h/r$, that is independent of radius.

The coordinates of the $x,y,z$ position (in coordinates
defined by the disk rotation axis) are then rotated
using $\chi$ and $\vartheta$
to account for the orientation of the system rotation axis with respect
to the viewer.
To produce each model image, 200,000 points in the disk are sampled.
Scattered starlight from these points 
are summed along each line of sight to produce a surface 
brightness image on the sky.
The brightness of each point in the disk depends on the assumed 
albedo times the normal optical depth of that portion of the disk 
multiplied by 
the flux from the star at that position.  The scattering amplitude
of the reflected light from each point in the 3D model was modified
by the Henyey-Greenstein scattering phase function. 
The scattering asymmetric parameter, $g$, describes the scattering
anisotropy ($g=0$ corresponds to isotropic scattering, $g=1$ to 
fully forward scattering).  However, we do not
vary $g$ but instead adopt a fixed value of $g=0.5$.  We chose
a representative value for $g$ because of the uncertainty
in the dust grain distribution and composition and because
of the broadness of the filter used for the {\it HST} observations.

We take into account absorption of starlight from the inner disk
by attenuating the starlight reaching the outer disk.
We use an axisymmetric attenuation function that depends on 
the spherical coordinate, $\theta$,  where $\sin \theta  = z/r$.
We assume that the opacity of the inner disk 
\begin{equation}
\tau_d(\theta) = \tau_0 \exp(-|\theta|/\theta_\tau)
\label{eqn:shadow}
\end{equation}
where $\tau_0$ is the opacity in the equatorial plane, and 
$\theta_\tau$ describes an angular scale length. 
Approximately 50\% of the total stellar
luminosity is absorbed by the inner disk.
An opaque torus blocking 50\% of the light would cover
angles $-30^\circ < \theta < 30^\circ$.
Because the star is an B9.5V star its spectrum is quite blue.
Consequently we expect $\tau_d \gtrsim 1$ at broad band optical
wavelengths for angles a few times smaller 
than the $\theta \sim 30^\circ$ 
required to block 50\% of the total stellar luminosity.

Because the disk surface brightness is  
observed to drop $\propto r^{-3}$ within $2.5''$ 
of the star \citep{grady,augereau01},
the disk normal optical depth must depend on radius.
Taking into account the $r^{-2}$ flux from
the star, we match the observed radial drop with a $-1$ radial power 
in the normal optical depth \citep{augereau01}.
The normal optical depth times the albedo is taken to be a 
power law function of radius, $\propto r^{-1}$.
We note that if there are large variations in the disk orientation 
then a different radial function may provide
a better fit to the surface brightness profile of scattered light.

\subsection{The precession angle}

If the warp is due to the tidal force of a planet interior to the disk, 
then the precession rate of the longitude of the ascending node 
is approximately 
\begin{equation}
\dot\Omega(r) \approx  -{3 \over 4}  n 
      \left({ M_p \over M_*}\right) 
      \left({ D \over r}\right)^2
\label{eqn:prec}
\end{equation}
as discussed by \citet{mouillet97}. This approximation is appropriate
for particles at low inclinations and in nearly circular orbits.
Here $n$ is the mean motion of the disk at radius $r$ (equivalent
to the angular rotation rate for a circular orbit),  $D$ 
is the semi-major axis of the planet, $M_p$ is the mass of the planet,
and $M_*$ is the mass of the central star.  
This precession rate is appropriate in the limit for $D \ll r$ and
is independent of the disk inclination.  This precession rate
also neglects the self-gravity of the disk and so is only appropriate
when the disk is low mass. 

After a time $\Delta t$, an initially flat disk will have 
a precession angle 
\begin{equation}
\Omega(r) = \Omega_0  + A_0
\left({ \left({r \over r_0}\right)^{-\beta} - 1}\right)
\label{eqn:Adef}
\end{equation}
where $\Omega_0$ is the precession angle at a reference radius, $r_0$.
When the precession is due solely to a single distant planet,
the constant 
\begin{equation}
A_0 =  
-{3   \sqrt{GM_*}  D^2 \Delta t \over 4 r_0^{\beta}}
      \left({ M_p \over M_*}\right) 
\label{eqn:Adef_planet}
\end{equation}
and $\beta = 3.5$, corresponding to precession angle
$\Omega \propto r^{-7/2}$.
However if there are multiple planets in the disk or
the disk itself contains contains mass, one may 
consider more general laws or power laws with $\beta < 3.5$.
In our numerical exploration we have explored variations
in $\Omega_0$, $\beta$ 
and $A_0$ to match the observed morphology of HD100546's disk.

If a planet internal to the disk is initially taken out
of the plane containing the disk then the tilt angle $i$ 
(with respect to the plane containing the planet) 
decreases 
with increasing radius; the situation considered by \citet{mouillet97}.
However if the disk itself is tilted via tidal forces from an external 
stellar encounter then $i$ would increase with increasing radius.
Consequently 
we allow the disk tilt angle, $i(r)$, to vary slowly with radius.  
At small radii where structure in the disk is difficult to resolve
in the presence of scattered light from the star, models
of the form given by Equation (\ref{eqn:Adef})
predict extremely tight corrugations.  
We let $i$ drop to zero at small radii, so that
tight corrugations in the inner region were removed
from the model images.  This allowed
us to keep a simple powerlaw form for the precession angle. 
We allowed the inclination to vary with radius smoothly by
using a spline function specified at 4 different radii  
$r=0,100,200$ and 400AU.   The inclination at $r=0$ was set 
to $0^\circ$.

Common parameters for models are listed in Table \ref{table:common}.
Parameters varied for individual models are 
listed in Table \ref{table:all}.
We first attempt to match the morphology of the disk with a geometric
model and then discuss physical models which can account
for the observed geometry.

\section{Model disks}

\begin{figure*}
\plotone{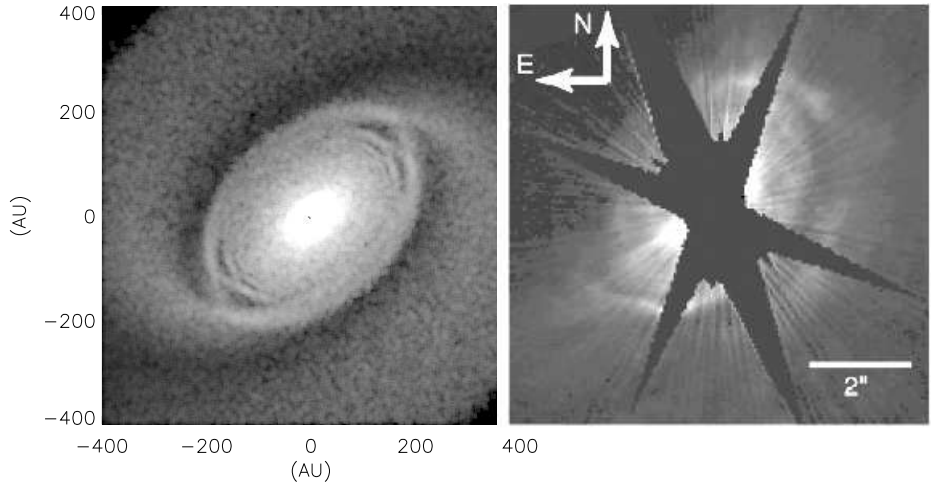}
\figcaption{
The left-hand panel shows the
log surface brightness of model warped disk, \#MA with
parameters are given in Tables \ref{table:common} and \ref{table:all}.
The middle panel shows log of the surface brightness of
the {\it HST} image of HD100546 from \citet{grady}.
Our model shows that a warped disk can exhibit
surface brightness variations in a two armed spiral pattern.
The surface brightness variations are due to the slope or
gradient changes in the disk surface with respect to the line
of sight and an equatorial shadow from an inner disk.
\label{fig:hdfig}
}
\end{figure*}

\begin{figure*}
\epsscale{0.70}
\plotone{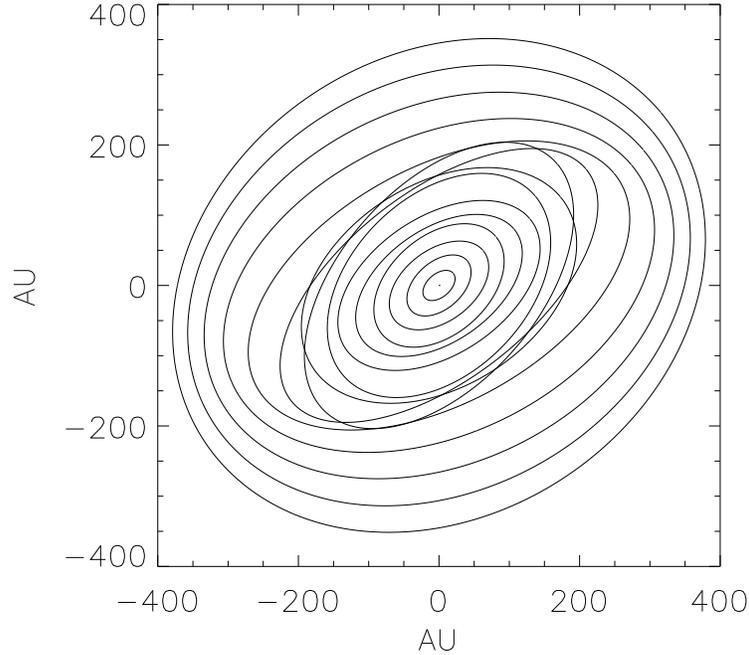}
\figcaption{
Projected circular disk rings at evenly spaced radii corresponding 
to the disk shown in figure \ref{fig:hdfig} and described
by model \#MA.
Regions where projected rings are in close proximity correspond to regions
of bright surface brightness (see figure \ref{fig:hdfig}). 
This is because these
regions have large slopes or surface gradients with respect
to the line of sight.
The equatorial shadow lies along the minor axes of the rings.
\label{fig:ring}
}
\end{figure*}

In Figure \ref{fig:hdfig} we show a model warped disk
in comparison to the {\it STIS} image HD100546 from Figure 1e by
\citet{grady}.    The parameters used to describe this
model are listed in Table \ref{table:common} and
as model \#MA in Table \ref{table:all}.
The rings comprising the disk are projected onto the
sky in Figure \ref{fig:ring}.
From a comparison between Figure \ref{fig:ring} and Figure \ref{fig:hdfig}
we see that locations where the rings are in close
proximity correspond to regions of higher surface brightness.  
These are regions where the disk slope or surface gradient
(with respect to the line of sight) is high.
A shadow from the inner disk is seen along the ring minor
axes.

\begin{figure*}
\plottwo{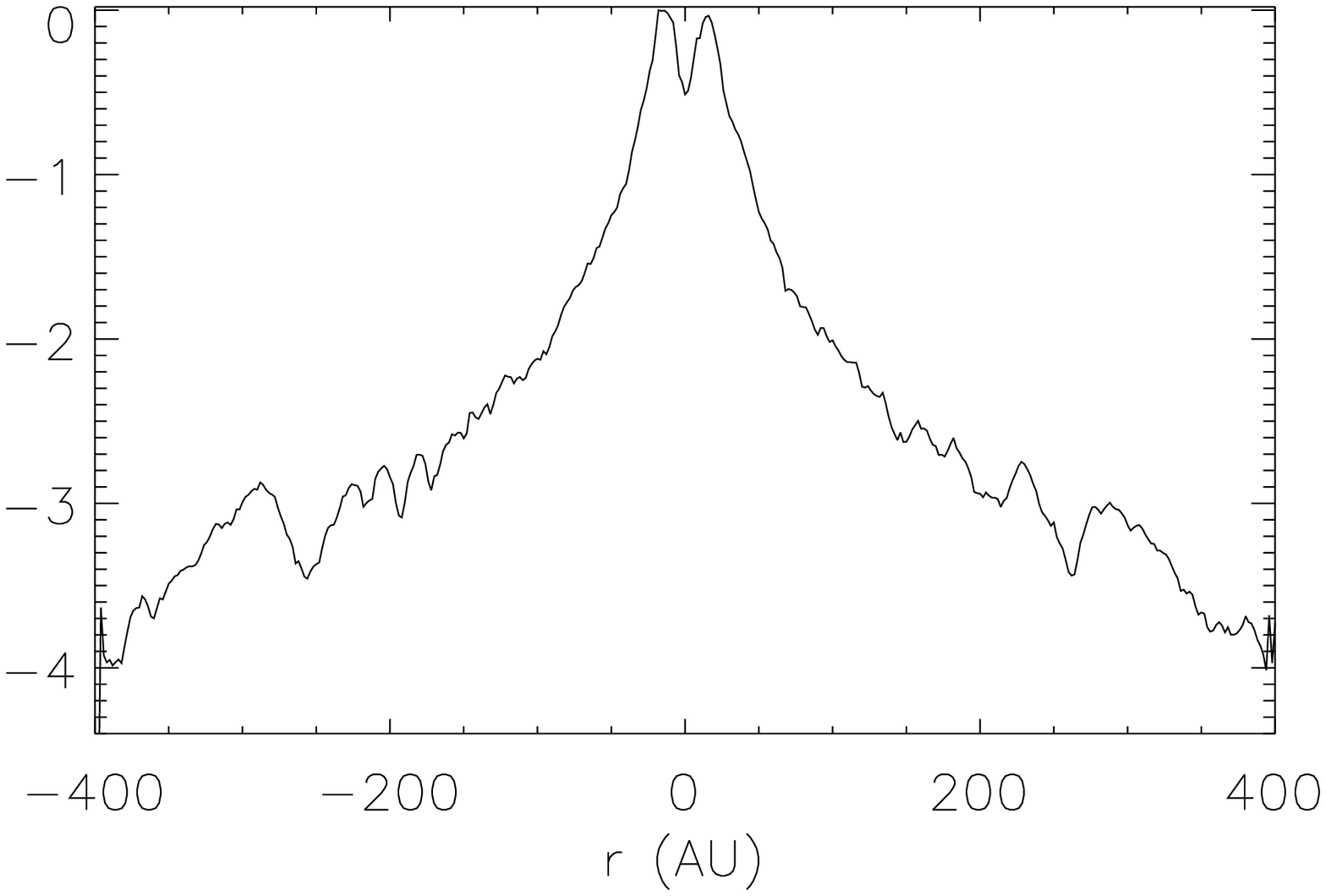}{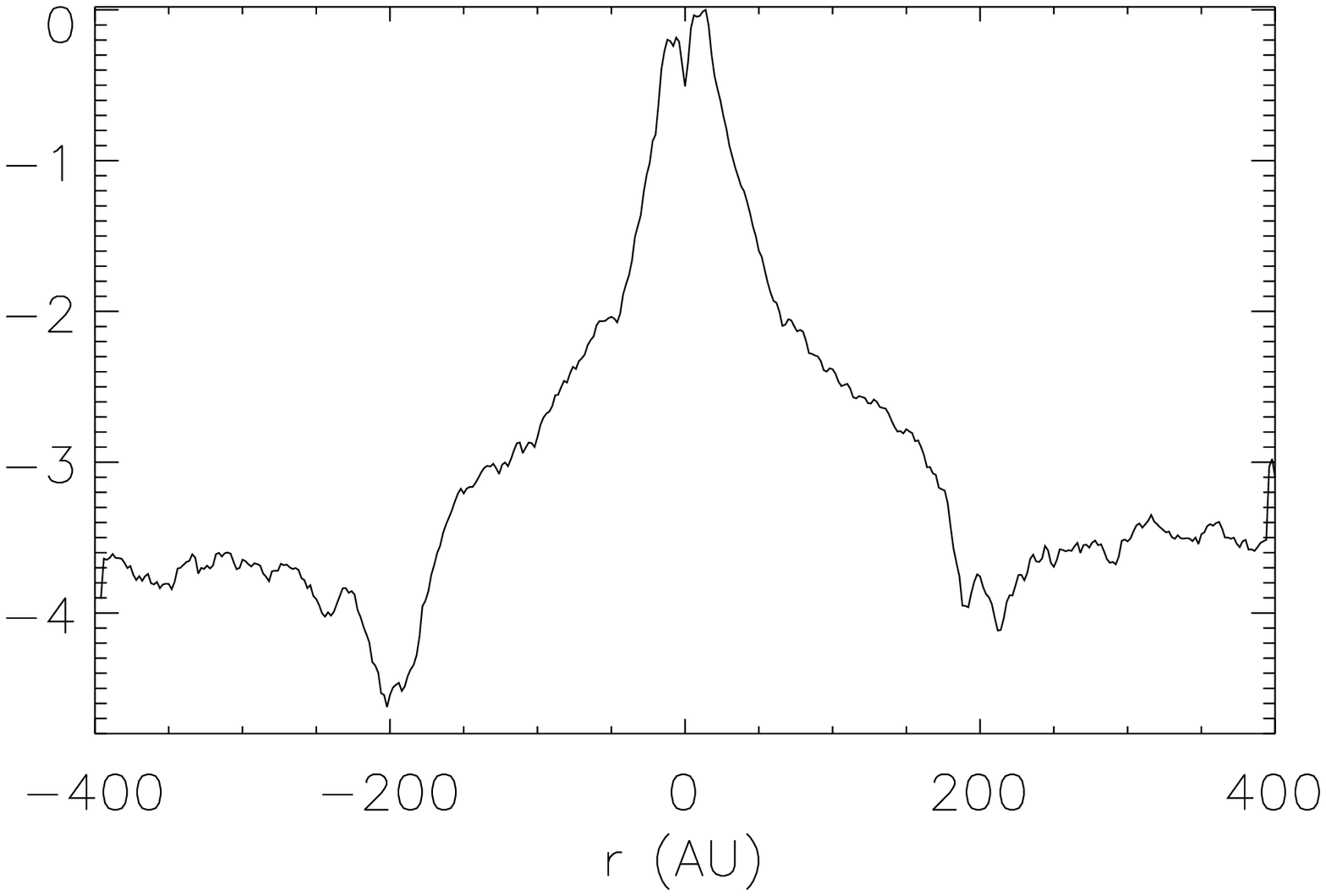}
\figcaption{
Major and minor axis surface brightness profiles for
model \#MA, that is also shown in Figures \ref{fig:hdfig} and \ref{fig:ring}.
Dips in the surface brightness profile are seen where there are
dark dustlanes due to equatorial shadowing by the inner disk.
Along the major axis at about 300 AU, an increase in the surface
brightness is seen at a fold in the disk.
A hole at $r = 13$AU has been placed near the star, primarily
to limit the flux range covered by the plot.
For the major axis profile, positive $r$ refers
to southeastern side of the disk.  
For the minor axis profile positive $r$ refers to the near or southwestern
side of the disk.  
The $y$-axis shows $\log_{10}$ of the surface brightness normalized
to the peak value.
\label{fig:slice}
}
\end{figure*}

Surface brightness profiles for Model \#MA along the major
and minor axes, at position angles $127$ and $37^\circ$,
are shown in Figure \ref{fig:slice}.
We compare our model surface brightness profiles with the major
axis profile shown in Figure 5a by \citet{grady} of the {\it STIS} image.
In both our model and the observed disk
one spiral feature corresponds to a bump in the major axis surface
brightness profile on the northwestern side 
at about 300AU from the nucleus.
The bump in the surface
brightness profile is about 0.2 in the log above a smoothly
dropping curve; approximately consistent with that seen in the
observed profile.
The increase in surface brightness corresponding to
the opposite spiral feature is much less prominent
on the southeastern side in both model and observed profile than
that on the northwestern side.
Dustlanes are not prominent in the major axis surface brightness 
profile (in both model and observed profile),  however they 
are deep in the model minor axis profile.
The dips in the major axis profile exhibited by our model 
in Figure \ref{fig:slice}a, corresponding to dustlanes, 
are deeper than those observed.  Equatorial shadowing in the model
is more extreme than that observed.  We discuss this problem 
in more detail below.
The southern side of the disk is brighter than the north side because
we allowed the scattering to be anisotropic 
and have taken the southern side to be nearer the viewer than the norther side.
Some anisotropy in the scattering is 
consistent with the excess surface brightness 
in the near-infrared seen on the 
southern side at $r \sim 3''$ reported by \citet{augereau01}.

We find that a warped model
such as model \#MA shown in Figures \ref{fig:hdfig}--\ref{fig:slice}
can provide a good explanation for the spiral features observed
in the disk of HD100546.  The warp causes apparent surface brightness
variations along two tightly wound 
spiral features, with darker regions within them.
The darker regions were described as lanes by \citet{grady}. 
Two effects can account for
regions of lower surface brightness: 
these regions are less inclined or more nearly perpendicular to
the viewer (more nearly face-on), or
they lie in the equatorial plane and so can be 
in the shadow of the inner disk.
Higher surface brightness regions are those
that rise above the midplane shadow.  They also
correspond to regions that have a steeper surface gradient 
or slope in the disk due to corrugations in the disk.
Low inclination evenly illuminated warps can exhibit
large spatial variations in the scattered light surface brightness,
however this only occurs if the disk is highly corrugated or twisted 
and the disk is optically thin.

In Figure \ref{fig:comp}, lower right-hand panel,
we show a nearly edge-on model \#ME, with parameters listed 
in Tables \ref{table:common} and \ref{table:all}.
This model is identical to model \#MA, shown in Figures 1-3
except the system is more highly inclined. 
Model \#ME can be compared to the edge-on disk
of Beta Pictorus.  Beta Pictorus is less twisted in its outer region 
compared to this model and does not have a shadow in its midplane.
The one planet model of \citet{mouillet97} would lead us
to expect that the disk is more highly twisted at smaller radii.
In future we may consider the possibility that
some of the unresolved structure in Beta Pictorus's or AU Microscopii's 
inner disks at visible
wavelengths might be explained with a warp that extends to smaller radii.
We note that
asymmetries between emission from one side of the disk compared to the opposite
side would be introduced because the optical depth could be high
at folds, and because of  scattering asymmetry.

\subsection{Sensitivity to parameters}

\begin{figure*}
\plotone{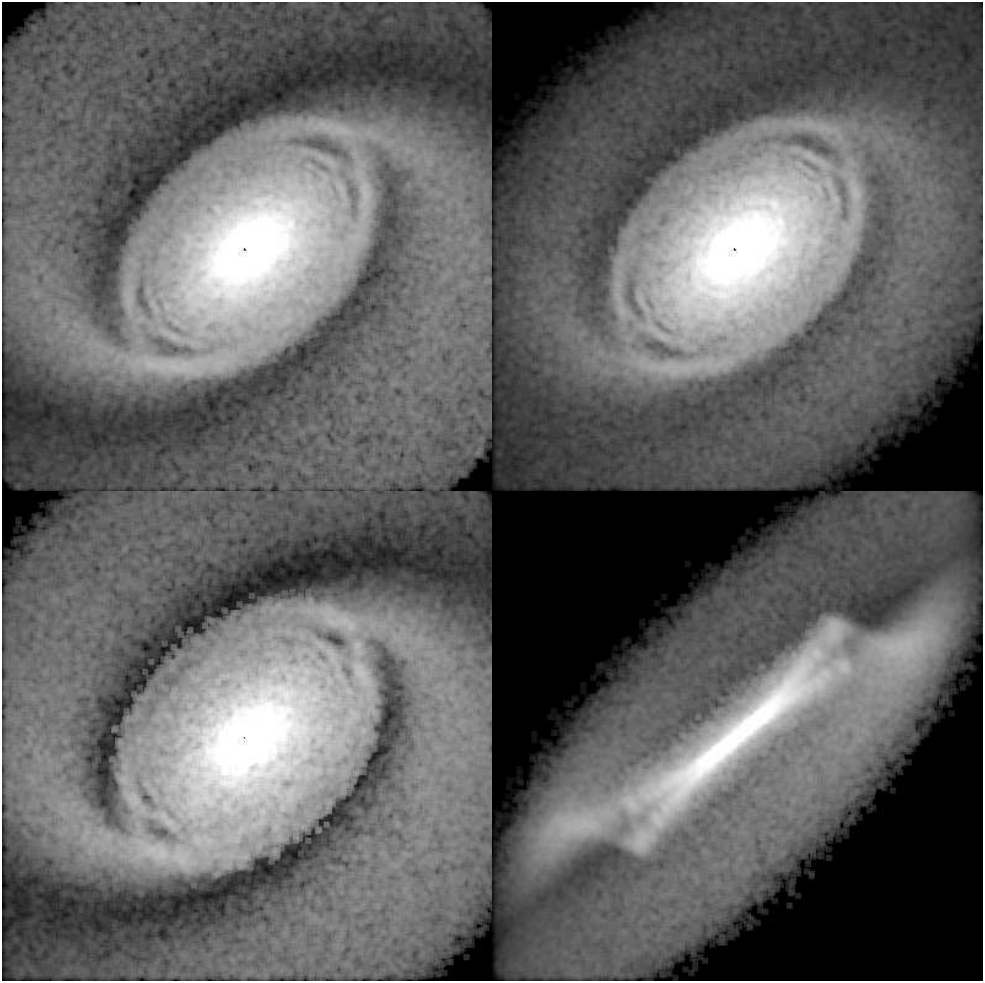}
\figcaption{
The effect of varying warped disk parameters. 
These models are similar to that (Model \#MA) 
shown in Figures \ref{fig:hdfig}, \ref{fig:ring}, and \ref{fig:slice}.
Varied parameters are listed in Table \ref{table:all}.
On the top left is Model \#MB which has a 
higher inclination at large radii than model \#MA.
The top right panel shows Model \#MC which has constant inclination
with radius.
The bottom left panel shows Model \#MD which has a higher equatorial
opacity for inner disk (larger $\tau_0$) than Model \#MA 
but a lower inner disk opacity angular scale height ($\theta_\tau$).
On the lower right we show a nearly edge-on disk, model
\#ME, to illustrate the extent of the twist, and for comparison
to edge-on systems such as Beta Pictorus.
\label{fig:comp}
}
\end{figure*}

In Figure \ref{fig:comp} we show the effect of varying some
of the parameters used to describe the disk.
Model \#MA has a disk inclination that increases with increasing
radius.  When the inclination is held constant, as is true in Model \#MC, 
shown on the top-right in Figure \ref{fig:comp}, the spiral arms
do not extend as far to the south-west and north-east as is seen 
in the scattered light image.  We find that 
the angular extent of each spiral feature is
smaller than that observed if $i$ is held fixed.
If the inclination with respect to the
system axis increases with radius, then the spiral features extend 
over a larger range of angles, as shown in Model \#MB on the upper left 
in Figure \ref{fig:comp}.  Model \#MB has an even higher outer inclination
than Model \#MA.
Disks with larger tilts (larger $i$) tend to produce higher surface brightness
variations in the spiral arms.  However if $i$ is increased past 10 degrees
then the disk can be folded with respect to the viewer and this reduces
the contrast of the spiral features.  Model \#MA and \#MB have
regions where multiple folds of the disk are encountered along the line
of sight (see Figure \ref{fig:ring}) and  only the fold edges are regions 
of high surface brightness.

It is interesting to note that the apparent ends of the two spiral arms
are about $180^\circ$ apart and oriented nearly 
along the major axis of the disk. 
Such a situation arises naturally from the warped models. 
In contrast, spiral arms
that are due to spiral density waves are not expected to end 
at locations $180^\circ$ apart.
The folds of the warped disk cause high surface gradient regions
(corresponding to higher surface brightness regions)
on either side of nucleus. However along the major axis these folds
are oriented along the line of sight. As a result, spiral features in
the model scattered light images tend to end along the disk major axis. 
Higher resolution models show a nested series of self-similar spiral features
inside the outermost ones.  This follows since
we have adopted a power law form for the precession angle.

Similar morphology to that observed 
is seen when the inner disk has a higher equatorial 
opacity $\tau_0$ but a shorter opacity angular 
scale length $\theta_\tau$.  Model \#MD 
shows such a model. This model has a slightly larger outer 
disk scale height which has the effect of smoothing the model 
surface brightness image.   Had we left the disk scale height similar
to that of model \#MA this model would have had extreme contrast in 
its dust lanes.
We find that thinner disks have more sharply defined spiral features and deeper 
dustlanes.   When the angular scale length of the equatorial shadow
is shorter, ($\theta_\tau$ is smaller),  the dustlane is more sharply
defined.  Higher equatorial inner disk opacity (large $\tau_0$) causes
deeper shadows.

As we commented above, we suspect that 
Model \#MA has deeper dustlanes than observed. The equatorial 
shadow for Model \#MA has an opacity of 1 (at visible wavelengths)
at equatorial angles $\theta =\pm 13^\circ$. This opacity is sufficiently
high that it is approximately consistent with the absorption of
$\sim 50$\% of the stellar light from the inner disk.
However this high opacity and large angular scale
length (in the inner disk opacity) 
implies that the outer disk is illuminated
by visible light that has been significantly 
attenuated by the inner disk; 
much of the outer disk has a tilt lower than $13^\circ$.
Our model computes the surface brightness at one wavelength only
corresponding to a  central optical  wavelength for the broad {\it STIS}
image.  It may be possible  to improve the model image
by integrating and summing images at different wavelengths.
It is also possible that wings of the stellar point spread function
have smoothed the appearance of the outer disk, reducing the 
actual surface brightness contrast.
The depth of the dustlanes in our model images can be decreased by 
increasing the angular scale length of the shadow, $\theta_\tau$.
However then the model midplane opacity must be reduced to reproduce
the observed morphology.  A midplane
opacity below 1 would be unrealistic as we expect
the midplane to be dense and optically thick at visible wavelengths.  
To improve the model we suspect that
we would require  a more complex function for the shadow than that
given by Equation (\ref{eqn:shadow}). This function
would necessarily be described by a larger number of 
parameters.  A more complex model is difficult to constrained with
the {\it STIS} image but could be constrained
by future high quality and multi-wavelength images.

The twistedness of the disk is set by $A_0$. 
Higher $|A_0|$ corresponds to more highly wound spiral features.
However, variations in $A_0$ can also 
change the radius of the spiral features if 
the radius $r_0$ is not simultaneously adjusted.
Because the spiral features are tightly wound we could
not place constraints on the parameter $\beta$ which
sets the dependence of $\alpha$  on radius, consequently
we set $\beta=3.5$, consistent with the model explored by
\citet{mouillet97}.
The contrast between the surface brightness on the near side
compared to that on more distant side is larger if the scattering
is more anisotropic; (the scattering asymmetry parameter $g$ is larger).

Based on our exploration of models with different parameters we have
found the following:
Only models with highly twisted disks have tightly
wound spiral features similar to those observed.  
Only models with inclination increasing with  radius have spiral
features that extended over a sufficiently large range of azimuthal
positions.  Only relatively thin disks, $h/r \lesssim 0.15$,
have
sufficiently sharp or fine features to be similar to the observations.
We find a degeneracy between the disk inclination, the angular form of
attenuation from the inner disk, and the disk thickness, because
these parameters all 
affect the contrast or amplitude of the spiral features.  
We did not find models with morphology similar to that observed with
disks higher than $i\sim 15^\circ$ and lower than $i\sim 6^\circ$. Low
inclination disks failed to exhibit sufficiently high surface brightness
variations and high inclination disks exhibited multiple folds along
the line of sight, reducing the surface brightness variations.
Models with small shadow angular scale lengths 
($\theta_\tau \lesssim 5^\circ$)
have dustlanes that are excessively deep, suggesting that
the upper layers of the inner disk have an
opacity distribution (as a function of $\theta$) with 
a moderately large angular scale length. 



\subsection{Mass constraints}

By matching the observed morphology
of the disk with our model, we can estimate the
extent that the disk is twisted.   This is described by 
parameters $A_0$, and $\beta$ (see Equation \ref{eqn:Adef}).
We consider here the hypothesis that the twist was caused
by an unseen inner planet that is misaligned with the outer disk 
(e.g., \citealt{mouillet97,augereau01b}).
We remind the reader that Equation (\ref{eqn:Adef_planet}) refers to
to a tilted disk with negligible mass which is perturbed by an inner planet.
The parameter $A_0$ depends on the time since the disk was initially tilted,
$\Delta t$,  and the mass and semi-major axis
of the hypothetical inner planet causing the precession.
We can assume that $\Delta t$ is less than the age of
the star or $\Delta t < 10^7$ years.  This allows
us to place a lower limit on the mass of the planet times
the square of the planet semi-major axis.
Using Equation (\ref{eqn:Adef}) and replacing $\Delta t$ with the age of the
system, $t_{age}$, we find
\begin{equation}
D^2 M_p  > {4 |A_0| M_* r_0^{7/2} \over 3 \sqrt{G M_*} t_{age}}.
\end{equation}
Computing these quantities for our value of $A_0$ (in radians) and reference
radius $r_0$ (listed in Table \ref{table:all}), we find
\begin{equation}
D^2 M_p \gtrsim 2.9 \times 10^4  AU^2 M_J
\end{equation}
where $M_J$ is the mass of Jupiter.
However,
this constraint is impossible to satisfy for a single Jovian mass at small
radius.    This constraint cannot be satisfied for a Jovian 
mass planet within the
lit edge of the disk at 13 AU \citet{grady05}.
A brown dwarf sized object at $D\sim 200$AU would open a gap
in the disk that would be observable in the images.
The simplest scenario of an initially misaligned low mass disk and
and a single inner planet fails to account for the twistedness of
the disk.

One way to account for the highly twisted disk would be if
the disk itself contained significant mass.
For example, a Jovian
mass of planetesimals between 100-200 AU but misaligned with the outer disk
could account for the twisted nature of the disk. 
If this mass is not confined to one body but extended then
the radial power of the precession angle would
be reduced.  This might account for the somewhat better match
of our model \#MB with $\beta = 2.5$ instead of 3.5.
HD100546's disk appears to be more twisted than 
Beta Pictorus's disk, however Beta Pictorus is 10-20 times older so 
the extent of the twist is less of a constraint on the planetary and
disk system. 

%


\subsection{Leading or trailing arms}

It may in future become possible to measure the sense of rotation 
of HD 100546's disk.  The disk rotation axis could be similar to
the star's rotation axis. 
Unfortunately the rotation axis of the star HD100546 is not known. 
\citet{clarke} attempted to measure
this axis from polarization measurements, however these 
were affected by scattering from the surrounding dust. 
In spite of this, we now discuss the sense of the warp in comparison
to the direction disk rotation.   Future measurements of either the disk
or stellar rotation could be used to support or refute dynamical
explanations for the spiral morphology. 

Equation (\ref{eqn:prec}) shows that
the inner disk should precesses faster than
outer disk and in the retrograde sense.  
Consequently we expect the disk to should twist
in the direction of rotation as the radius
increases.  If we differentiate Equation (\ref{eqn:prec})
we find that $d\Omega/dr$ is positive and so 
$\Omega$ increases in the direction of rotation.  
The spiral features would
be {\bf leading} instead of trailing.  
If the twist is a result of precession induced by bodies
in the disk interior, we 
would predict that the disk is rotating 
clockwise on the sky. We expect the northwestern side
to be blueshifted from the system center of mass, 
and the southeastern side to be redshifted.  
The southwestern side is probably closer to
us, based on the nebulosity on the southern side seen in the {\it STIS} image
\citet{grady} and because the excess seen in the scattered
near-infrared light at $r\sim 3''$ on the southern side
\citep{augereau01}.  

\section{Summary and Discussion}

In this paper we have presented a new model for the spiral features 
seen in visible scattered light images of HD100546's disk.
We reproduce the two-armed spiral features with a highly twisted
warped disk model.  The disk inclination with respect
to the systems ecliptic is low, 6 to 15$^\circ$ degrees 
and increasing with radius.  
The disk is fairly thin with a ratio of FWHM/$r$ 
of $\sim 0.15$.   Surface brightness 
variations are due to a high surface gradient with respect to the line of
sight in the folds of the disk.  Dark lanes increase the contrast 
of the spiral features and are caused by equatorial 
shadowing from an inner unresolved disk.
As a nearly Keplerian system can maintain a warp for long
periods of time (secular timescales rather than rotational periods),
this model has some advantages over transient spiral models.
The observed spiral features end at positions approximately
located on the disk major axis (as seen on the sky).  
This is a feature common to moderate inclination 
warped disk models that would not
be exhibited by a spiral density wave model for the
spiral features.

However, the morphology requires the disk to be so twisted
that the tidal force from a Jovian mass object in the inner
disk clearing cannot have
induced the twist during the lifetime of the star.
One possibility is that
the disk warp could have been induced by a significant mass
(Jovian mass) of objects that are inclined with respect to the outer disk, 
well outside the inner clearing, at intermediate
radii between 100-200 AU.  

We now discuss proposed mechanisms for accounting for the tilt
between the disk and objects in the inner disk.
A stellar encounter could tidally induce
a tilt in the outer disk.  In this case we would expect the 
disk tilt or inclination would increase with radius 
(e.g., as calculated by \citealt{kobayashi}).
Our models better match the angular extension of the spiral
feature when the 
tilt increases with radius, suggesting that this may
be the case for HD100546's disk. 
Such a stellar encounter, while unlikely in the field, could have
been probable in a denser stellar environment such
as the star's birth cluster.
An alternative scenario is that resonances between multiple planets
cause an inclination increase in one of the planets \citep{thommes03}.
In this case, the disk would have tilt or inclination 
that decreases with radius \citep{mouillet97}.    
Also this model would likely require vertical
resonances for planets at large semi-major axes to account for
the highly twisted disk.
The different dynamical scenarios predict
different radial forms for the angles
used to describe the warp, $i$ and $\Omega$.
Consequently better observations and accompanied modeling of the HD100546's 
disk and other circumstellar disks will produce better
constraints on the disk tilt and precession angle as a function
of radius.   
These in turn will allow better tests of the dynamical models 
for the warp formation and evolution.  The 
dynamical models have the capability
of providing unique constraints on the mass distribution 
in the outer disk.

In this paper we have used a very simple model to 
take into account equatorial shadowing due to the
opacity of the inner disk.  
Unfortunately we find a degeneracy in our models
between the disk tilt, the angular form of
attenuation from the inner disk, and the disk thickness. 
This is because these parameters all 
affect the surface brightness contrast or amplitude of the spiral features.  
Shadowing is likely to be a strong function of color, consequently 
multi-color imaging 
may be able to probe the structure of the inner unresolved disk
as well as better constrain the geometry and structure of the outer disk.
Warped disks seen at visible wavelengths should exhibit asymmetries
due to optical depth variations, 
and asymmetries due to forward scattering.
Both of these affects should be more severe at bluer wavelengths, 
providing a possible way to discriminate between geometrical models.
We note that some asymmetries and perceived clumps in visible light
in edge-on warped disks (e.g., as seen in AU Microscopii; \citealt{liu}), 
might in future be explained with a warped disk rather than 
a model containing eccentric rings.
However asymmetries observed in thermal or mid-infrared emission,
such as is observed in Beta Pictorus by
\citet{telesco}, could not be explained via a warped disk model as
at these wavelengths 
we expect the disks to be more nearly optically thin 
the emission should be nearly isotropic.

The observations of \citet{eisner} suggest that
HeBe stars are not highly warped.  However, moderate inclination warps, such
as found here would not violate the comparison of submillimeter and
near-infrared position angles and ellipticities.
It is possible that highly tilted warped disks may provide 
alternate explanations
for excess far-infrared emission in some systems 
as such disks cover a larger solid angle and so can absorb more stellar 
flux at large radii than a coplanar disk of the same thickness.

\acknowledgments

I thank the Research School of Astronomy and Astrophysics of the 
Australian National University 
and Mount Stromlo Observatory 
for hospitality and support during Spring 2005.
Support for this work was in part
provided by National Science Foundation grant AST-0406823,
and the National Aeronautics and Space Administration
under Grant No.~NNG04GM12G issued through 
the Origins of Solar Systems Program.  
Support was also provided by
the National Science Foundation to the Kavli Institute
for Theoretical Physics under Grant No.~PHY99-07949.

{}

\clearpage

\begin{deluxetable}{lc}
\tablewidth{0pt}
\tablecaption{Common Parameters describing the warp models \label{table:common}}
\tablehead{
\colhead{Parameter}    &
\colhead{Value}    
}
\startdata
$\chi$   & $40^\circ$   \\ 
$A_0$    &$-260^\circ$  \\
$r_0$    & 250 AU       \\
$\Omega_0$  & $230^\circ$ \\
$\beta$    & 3.5    \\
\enddata
\tablecomments{
The models shown in Figures \ref{fig:hdfig}, \ref{fig:ring},
\ref{fig:slice} and \ref{fig:comp} have these parameters in common.
The position angle of the system rotation axis on the sky is denoted by $\chi$. 
The precession angle at the reference radius $r_0$ is $\Omega_0$.
The reference radius $r_0$ is given in AU (100AU $\sim 1''$ on the sky
for HD100546).
The parameters $A_0$ and $\beta$ describe the sensitivity of the precession
angle $\Omega$ with radius or the twistedness
of the warp; see Equation (\ref{eqn:Adef}).
The angles $\chi$, $\Omega_0$ and $A_0$ are given in degrees.
}
\end{deluxetable}

\begin{deluxetable}{lccccc}
\tablewidth{0pt}
\tablecaption{Parameters describing the individual models \label{table:all}}
\tablehead{
\colhead{}    &
\colhead{\#MA}   &
\colhead{\#MB}   &
\colhead{\#MC}   &
\colhead{\#MD}   &
\colhead{\#ME}   
}
\startdata
$\vartheta$ 
    & $52^\circ$    & $52^\circ$   & $52^\circ$  & $52^\circ$    & $89^\circ$ \\
$i$         
    &$3,6,15^\circ$ &$3,5,20^\circ$&$6,6,6^\circ$& $3,6,15^\circ$& $3,6,15^\circ$   \\
$h/r$       
    & 0.05          &  0.05        & 0.05        &  0.08         & 0.05 \\
$\tau_0$    
    &    3.0        &  3.0         & 3.0         &  5.0          & 3.0\\
$\theta_\tau$
    &$12.0^\circ$  & $12.0^\circ$ &$12.0^\circ$ & $5.0^\circ$   & $12^\circ$\\
\enddata
\tablecomments{
The inclination of the system rotation axis with respect to the viewer is $\vartheta$.
The inclination of the warped disk with respect to the system
rotation axis is $i$.
The angles $\vartheta$, and $i$  are given in degrees.
The disk aspect ratio, assumed to be constant with radius is denoted
by $h/r$. 
The standard deviation of the vertical density distribution
is $h$ and the FWHM is $2.35 h $.
Inclination angles are given at three radii, $r=100$, 200 and 400AU, and
we set $i=0$ at $r=0$.
A spline function was fit between these values.
The opacity $\tau_0$ describes the opacity of the inner disk in
the equatorial plane.  The parameter $\theta_\tau$ describes the angular
scale height of the inner disk's opacity in degrees.
Surface brightness profiles, projected rings and images 
for model \#MA are shown in Figures \ref{fig:hdfig}, \ref{fig:ring},
\ref{fig:slice}. Images are shown in Figure \ref{fig:comp} for Models 
\#MB--\#ME.
} 
\end{deluxetable}

\end{document}